\documentclass[%
 reprint,
superscriptaddress,
 amsmath,amssymb,
 aps,
pra,
]{revtex4-2}
\usepackage{comment}
\usepackage{wasysym} 
\usepackage{tikz}

\usepackage{xr}
\usepackage{ dsfont }
\usepackage{comment}
\usepackage[normalem]{ulem}
\usepackage{soul}
\usepackage{lineno}

\usepackage{bm}

\usepackage{graphicx}

\usepackage{color}

\newcommand{\red}[1]{\textcolor{black}{#1}}

\begin{document}

\preprint{APS/123-QED}

\title{
Entropic Clustering of Stickers Induces Aging in Biocondensates}

\author{Hugo Le Roy}
\email{HCV.Le-Roy@proton.me}
\affiliation{Institute of Physics, \'Ecole Polytechnique F\'ed\'erale de Lausanne---EPFL, 1015 Lausanne, Switzerland}

\author{Paolo De Los Rios}
\affiliation{Institute of Physics, \'Ecole Polytechnique F\'ed\'erale de Lausanne---EPFL, 1015 Lausanne, Switzerland}

\begin{abstract}
Biomolecular condensates are cellular
phase-separated droplets that usually exhibit a viscoelastic mechanical response. A behavior rationalized by modeling the complex molecules that make up a condensate as stickers and spacers, which assemble into a network-like structure. 
Condensates usually exhibit a solidification over a long period of time (days), a phenomenon described as aging. The emergence of such a long timescale of evolution from microscopic processes, as well as the associated microscopic reorganization leading to aging, remains mostly an open question. In this article, we explore the connection between the mechanical properties of the condensates and their microscopic structure. We propose a minimal model for the dynamic of stickers and spacers, and show that entropy maximization of spacers leads to an attractive force between stickers. Our system displays a surprisingly slow relaxation toward equilibrium, reminiscent of glassy systems and consistent with the liquid-to-solid transition observed. To explain this behavior, we study the clustering dynamic of stickers and successfully explain the origin of glassy relaxation.
\end{abstract}

\maketitle

\section*{Introduction}
Liquid-liquid phase separation has emerged as a major organizing principle for the compartmentalization of cellular processes into membraneless organelles known as biomolecular condensates. Condensates are associated with various functions of the cell, although their precise role often remains elusive \cite{bananiBiomolecularCondensatesOrganizers2017}, to the point that it is not even clear if some of them are functional, incidental or pathological \cite{shinLiquidPhaseCondensation2017}. Similarly, little is known about the complete composition of condensates; however, one or several scaffolding components that drive the phase separation can usually be identified, including proteins, DNA \cite{ryuBridginginducedPhaseSeparation2021,rippeLiquidLiquidPhase2022}, or RNA \cite{hondeleDEADboxATPasesAre2019,aulasStressspecificDifferencesAssembly2017}. Despite their diversity, condensates also display common features. It is usually accepted that their formation requires weak multivalent interactions \cite{leeRecruitmentMRNAsGranules2020}, mediated by disordered portions of proteins \cite{boeynaemsProteinPhaseSeparation2018} or by RNA/DNA-protein interactions \cite{brackleyNonspecificBridginginducedAttraction2013}. Similarly, micro-rheology experiments suggest that biocondensates exhibit a general viscoelastic mechanical response \cite{jawerthSaltDependentRheologySurface2018,alshareedahProgrammableViscoelasticityProteinRNA2021}. This picture has been rationalized through the sticker and spacer model, which describes proteins as generic sequences of sticky (stickers) and non-sticky (spacers) regions \cite{choiPhysicalPrinciplesUnderlying2020}. 
The complex interactions that make up a condensate give rise to an emergent long-time transition to a solid-like state,
usually described as aging \cite{jawerthProteinCondensatesAging2020,linsenmeierDynamicArrestAging2022}. This process has been associated with protein denaturation \cite{faragCondensatesFormedPrionlike2022}, amyloid formation \cite{rayASynucleinAggregationNucleates2020} and, from a physiological perspective, with neurodegenerative diseases \cite{shenLiquidtosolidTransitionFUS2023}.
Current physical pictures of condensate aging either rely on the modeling of specific proteins' interaction -- like $\beta$-sheet stacking \cite{ranganathanPhysicsLiquidtosolidTransitions2022} --, or do not model the microscopic dynamic of the components \cite{takakiTheoryRheologyAging2023,garaizarAgingCanTransform2022}. Yet, aging seems to be a fairly general behavior, {\color{black}
suggesting that its occurrence does not depend on specific compositional details.
Together with experimental evidences of a disordered-to-order transition \cite{alshareedahSequencespecificInteractionsDetermine2024} or the formation of substructures \cite{faragCondensatesFormedPrionlike2022,wuSingleFluorogenImaging2023} suggests that some of the features of this transition can be governed by a universal physical mechanism.}
A general microscopic picture explaining how large timescales (days or more) emerge from a collection of microscopic processes, that typically occurs on much shorter timescales, ($\mu s$) is missing.

In this work, we use the sticker and spacer framework to model the microscopic structural rearrangement taking place in condensates, and leading to aging. 
{\color{black}We do not investigate the phase separation process, instead, we consider the components of a droplet as a heterogeneous collection of localized, attractive regions (stickers) and longer, neutral regions (spacers).}
Stickers reversibly associate via non-covalent interactions, while spacers provide flexibility and connection between stickers, resulting in the formation of a percolating network. The ensuing mechanical behavior of this network is typical of an associative gel, characterized by a viscoelastic response to stress. At {\color{black}timescales much smaller than the sticker lifetime,}
the gel displays solid, elastic-like behavior due to the entropic elasticity of the spacers, resisting deformation. Conversely, over longer timescales, the reversible binding and unbinding dynamic of stickers enable the material to flow, displaying liquid-like characteristics.
Homogeneous associative gels display a soft relaxation over a single characteristic timescale, directly related to the unbinding dynamics of the stickers \cite{albertoparadaIdealReversiblePolymer2018}. 
In contrast, heterogeneous gels feature multiple coexisting timescales, each corresponding to a different sub-region within the system, and a complex relaxation behavior emerges from their superposition \cite{songNonMaxwellianViscoelasticStress2023}.
Strong heterogeneity results in a broad distribution of timescales, sometime heavy tailed. In these extreme cases, a portion of these timescales are arbitrarely large and relaxation seemingly never ends, a phenomena typical of glasses, and known as aging \cite{lielegSlowDynamicsInternal2011, berthierDynamicHeterogeneityAmorphous2011}.
To investigate how aging can emerge from the collective rearrangement of the gel components, we model them as stickers and spacers. 
the former acting as diffusing particles that can bind to (and unbind from)  the latter, that represent Gaussian polymers, possibly capturing intrinsically disordered regions of proteins or long RNA strands.
Using a combination of analytical results and  simulations, we show that an entropic Casimir-like force drives the clustering of stickers, reminiscent of substructures experimentally observed \cite{faragCondensatesFormedPrionlike2022, ryuBridginginducedPhaseSeparation2021,darBiomolecularCondensatesForm2024}. 
We find that collective rearrangement of the stickers induce a slow-down of the dynamic of the system. As a consequence, the relaxation toward equilibrium is surprisingly slow and reminiscent of the usual glass dynamic \cite{gotzeLogarithmicRelaxationGlassforming2002}.
To explain our observations, we introduce a minimal model for the dynamic of clusters, which provide a scaling law for the dynamical slowing-down.
{\color{black}Finally, we compute the associated dynamic modulus of our system which demonstrate the long-time solidification resulting from aging.}
Despite its simplicity, our model displays a rich phenomenology leading to glassy dynamic which includes emergence of large timescales, ergodicity breaking and aging.
{\color{black}By using a minimal model for the emergence of soft glassy relaxation, we propose a fairly general driving mechanism for the aging of biocondensates.}


\section*{Results}
\subsection*{Dynamic of stickers and spacers}


We consider a situation in which the system is already demixed, and focus on the dynamics of stickers and spacers. Spacers—corresponding to RNA/DNA strands or intrinsically disordered segments of proteins—are modeled here as linear Gaussian polymers. This description assumes a regime where topological constraints and excluded volume interactions are negligible. 

\red{This minimal model ensures that the system remains analytically tractable and
numerically accessible over long timescales, and it allows 
us to highlight the emergence of entropic attractive
forces driving between stickers, even if their magnitude might
be quantitatively modulated by steric repulsion in a more accurate description.}

\begin{figure}[!ht]
\centering
\includegraphics{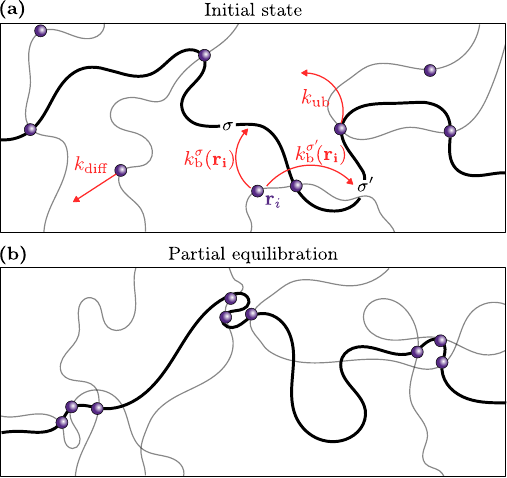}
\caption{Schematic representation of our sticker and spacer model. The disordered background potential is represented as a color coding, while stickers are represented as diffusing particles. We focus on a single polymer dynamic highlighted in bold, while the other effective polymers are in shaded. \textbf{(a)} Represents the initial state in which stickers are homogeneously spread in space. The dynamics of stickers are also represented as small arrows: stickers bound to one polymer can diffuse or bind to another, whereas those bound to two polymers create a cross-link and must unbind before diffusing further. \textbf{(b)} Qualitatively represent a typical final state as predicted by our model.
}
\label{fig:2}
\end{figure}

Stickers are modeled as diffusive particles that can transiently bind to spacers, forming reversible cross-links. Stickers may represent adhesive domains within a polymer or separate binding proteins. We model both cases similarly by simulating a single polymer \red{\sout{strand}} while treating the surrounding environment, composed of other polymers and stickers, as an effective bath of mobile particles that diffuse freely when unbound but become immobilized when bound to the simulated polymer.
\red{Specifically, we focus on a single polymer chain interacting with multiple linkers. We define a "polymer strand" as any segment of the chain located between two consecutive binding sites.}
This dynamic for stickers reflects the formation of cross-links with other effective spacers that impair their diffusion and locally restrict polymer fluctuations as shown in Fig.~\ref{fig:2}\textbf{(a)}.
Assuming that the dynamic of the one polymer simulated is representative of the others in the system, we are able to compute the viscoelastic response of such a polymer network.

While the diffusion dynamics of stickers may differ from that of free particles depending on their exact nature, we will show that the long-time evolution of the system is controlled by the binding and unbinding dynamics. As a result, we expect that variations in the diffusion properties will not significantly affect our results.

Here, we neglect the possibility that a sticker simultaneously interacts with other effectif spacers. This may lead to a slight overestimation of the number of free diffusing stickers or an underestimation of their effective diffusion.
The goal of this modeling is to describe the essential physical trade-off: cross-linking lowers the system’s free energy but constrains its ability to rearrange

\red{To describe the dynamic of the stickers, we now consider the probability of the $i$-th sticker to be in position $\bm{r}_i$ at time $t$ in the bound state with any polymer strand denoted by $\red{p_\text{b}(\bm{r}_i,t)}$}. Similarly, we write the probability to be unbound $p_\text{ub}(\bm{r}_i,t)$.
Given the previously introduced dynamics, the time evolution of these two probability distributions is governed by the two coupled master equations:
\begin{equation}
\begin{aligned}
\red{\partial_t p_\text{b}(\bm{r}_i ,t) =}&\red{ \sum_\sigma k_\text{b}^\sigma(\bm{r}_i) p_\text{ub}(\bm{r}_i,t) - k_{ub} p_\text{b}(\bm{r}_i,t)} \\
\red{\partial_t p_\text{ub}(\bm{r}_i,t) =}&\red{ k_{ub}p_\text{b}(\bm{r}_i,t) - \sum_\sigma k_\text{b}^\sigma(\bm{r}_i,\ell) p_\text{ub}(\bm{r}_i,t)} \\ &\red{+ D\nabla^2 p_{ub}(\bm{r}_i,t),}
\end{aligned}
\label{eq:master_equation}
\end{equation}
\red{where $k_\text{b}^\sigma(\bm{r_i})$ is the binding rate of the linker located in $\bm{r_i}$ with the polymer strand $\sigma$.}
We first write the unbinding rate using a single energy scale under the Kramers approximation:
\begin{equation}
k_\text{ub} = \frac{1}{\tau_0} e^{-\beta E_\text{b}},
\label{eq:unbinding_rate}
\end{equation}
where $\beta = 1/(k_BT)$ is the inverse temperature, $E_b$ the binding energy scale, and $\tau_0$ a timescale associated with the microscopic escape process. While the complexity of real proteins' interaction should lead to a distribution of binding energies, we show here that this assumption is not necessary for the system to display a glassy dynamic, and address the effect of a more complex model in the discussion section.

Concerning the binding rate, we assumte that polymers have time to equilibrate between two binding/unbinding events. As a consequence, we represent the polymers through their equilibrium spatial probability distribution.
We show in Supplementary Material that the probability for a given conformation of the polymer to meet a sticker located at $\bm{r}$ at a lineic position $\ell$ is given by:
\begin{equation}
\red{P_\text{meet}(\bm{r}) = \int \text{d}\ell e^{S_\text{b}(\bm{r},\ell) - S_\text{ub}} b^2}.
\label{eq:Pmeet}
\end{equation}
where $S_\text{b}$ and $S_\text{ub}$ correspond to the entropy of the polymer strand when bound and unbound to the sticker, respectively. \red{$b$ is the typical size of a sticker, taken equal to $1$ in the rest of the paper.}
Our approximation of fast polymer conformational exploration means that the limiting timescale is the binding time. Thus, we write the binding rate of the sticker to the polymer as:
\begin{equation}
k_\text{b}(\bm{r}) = 1/\tau_0  P_\text{meet}(\bm{r})
\label{eq:binding_rate}
\end{equation}
Together with (\ref{eq:unbinding_rate}),  this expression respects detailed balance, thus guaranteeing that the evolution of the system is consistent with thermodynamic equilibrium. 
Notice that $S_\text{b}(\bm{r},\ell)$ introduces a coupling between the dynamic of the different stickers ultimately depends on where all the other stickers are bound (see Supplementary Material for the full expression). As a result, we cannot directly solve this system of equation without further approximation.
In the next section, we use a Gillespie simulation to study the time evolution of our system.

\subsection*{Early time behavior}
\begin{figure}
\centering
\includegraphics[width = 0.45\textwidth]{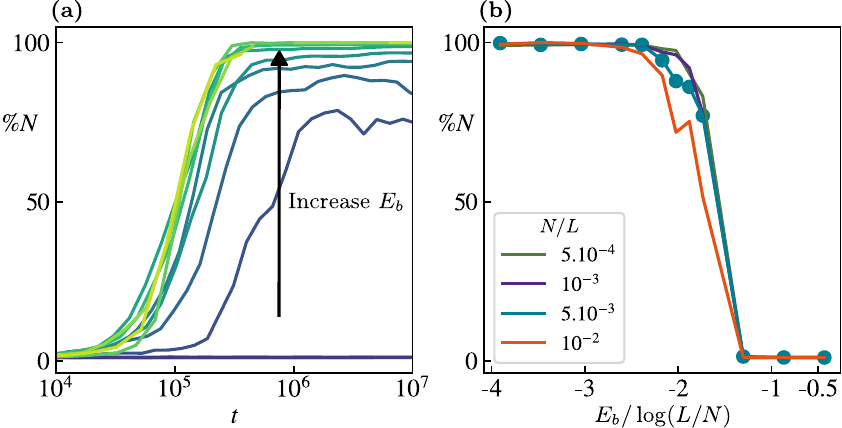}
\caption{\textbf{(a)} Early time evolution of the Percentage of bound stickers in the system. Here, we use $L = 2.10^4$ and $N=100$. \textbf{(b)} Plateau value of the percentage of bound stickers as a function of the binding energy, rescaled by the lineic density of stickers. The scattered points correspond to the measurements obtained from the panel \textbf{(b)}.}
\label{fig:Nlinkers}
\end{figure}
We simulate a polymer of length $L$, which defines the system size, surrounded by $N$ initially unbound stickers spread homogeneously in a 3D space.
Each Gillespie step consists in a move chosen among sticker diffusion, binding to a polymer in the vicinity of the sticker, or unbinding if it was bounded.
\red{All the rates are computed at the beginning of the step according to Eqs.~(\ref{eq:binding_rate}) and (\ref{eq:unbinding_rate}) using Eq.~(S5) of the supplementary material, and a move is selected with a probability proportional to its rate.} After selecting, and applying a specific move, a time increment is drawn from an exponential distribution with rate equal to the sum of rates of all the possible moves of the system.
After each move, the rates are locally updated consistently with the new configuration of the system.

Looking at the early time evolution of the system, we observe in Fig.~\ref{fig:Nlinkers}~\textbf{(a)} a rapid increase in the number of stickers bound to the polymer until a plateau is reached. We show in Fig.~\ref{fig:Nlinkers}~\textbf{(b)} how the height of the plateau depends on the binding energy, observing a sharp increase from $0\%$ of the bound stickers to $100\%$ at a critical value of the binding energy. This phenomenon is reminiscent of the phase transition observed in the Poland-Scheraga model of DNA denaturation \cite{polandPhaseTransitionsOne1966}. In their model, two DNA strands can bind to one another through sticky base pairs that are regularly spaced along the strands. Similarly to our model, the energetic binding energy must overcome the loss of entropy associated with the binding. As the linear density of stickers increases, it becomes increasingly favorable to bind them as the entropic cost for successive binding decreases. Their theory predicts a phase transition with a critical energy scaling, $E_c \propto \log(L/N)$. We observe that our model consistently displays a similar scaling based on the collapsed curves of Fig.\ref{fig:Nlinkers}
As a consequence, for our system, low values of the binding energy (or high temperature) mean that stickers are mostly unbound, and the system mechanically responds like a liquid. If the value of the binding energy is higher than its critical value (or the temperature lower than its critical value), stickers remain bound most of the time, leading to a viscoelastic mechanical response similar to that of an associative gel. This type of gel displays a solid-like mechanical response at timescales lower than the unbinding time and liquid-like at timescales much larger than the unbinding time. In this work, we focus on the viscoelastic aging regime; for this reason, we always consider the binding energy above its critical value. In this case, unbinding becomes the time-limiting process for the dynamics of the system.
In the following, we do not discuss the influence of the binding energy, which can be ruled out by rescaling times with respect to $1/k_\text{ub}$. In the high energy regime, the phenomena discussed are insensitive to the value of the binding energy.
\subsection*{Glassy relaxation}
\begin{figure}
\centering
\includegraphics[width=0.45\textwidth]{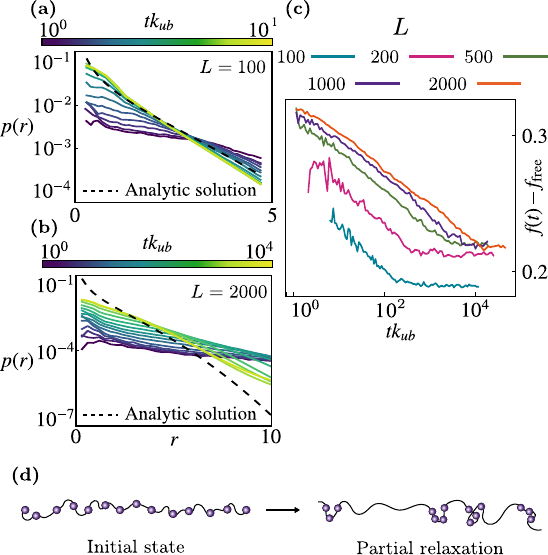}
\caption{\textbf{(a)} sticker distance probability distribution at different time-point. The color coding corresponds to the time, and the analytical solution is provided as dotted line. We simulate systems with fixed lineic density: $N/L = 5.10^{-2}$.
Small systems of length $L=100$ successfully converge toward the predicted equilibrium distribution. \textbf{(b)} Large systems, of length $L=2000$, evolves toward the analytic solution without reaching it.  \textbf{(c)} We follow the equilibration process by looking at the decay of the free energy per unit length over time, the initial binding regime as been removed for improved visibility. Using a fixed lineic density of stickers, we expect all curves to converge to the same relative value. However, large systems are stuck at a higher value of the free energy, stressing its inability to reach equilibrium.
\textbf{(d)} Schematic representation of the partial equilibration of the system into sub-clusters.
}
\label{fig:time_evolution}
\end{figure}
Once all stickers are bound, the system's energy becomes essentially fixed. However, the entropy of the polymer also depends on the positions of the stickers. We study the equilibration of the system through the diffusion of stickers. 
We first derive a mean field model of the equilibrium pair probability distribution for two stickers.
\red{To obtain an analytical form, we employ the single-ended propagator approximation (see Supplementary Material), which is justified here because the probability density is highly peaked near the tethering points (}$\red{r\approx R_l , R_r}$\red{) where this approximation holds.}
We find that for a system with a density of stickers $L/N$, the probability of finding two stickers at a distance $|\bm{r}|$ from each other is given by: 
\begin{equation}
p(\bm{r}) = \frac{3\text{ erfc}\left(\sqrt{\frac{3}{2aL/N}}|\bm{r}|\right)}{2 \pi a L/N |\bm{r}|}
\label{eq:pcf}
\end{equation}

The simulated time evolution of the probability distribution toward the expected equilibrium analytic solution for smaller systems is depicted in
Fig.~\ref{fig:time_evolution}~\textbf{(a)}.
Contrastingly, Fig.~\ref{fig:time_evolution}~\textbf{(b)} shows that larger systems, while converging similarly, do not reach our approximated equilibrium solution.
\red{The time evolution of the pair probability distribution reveals a contraction of the system. We compute the corresponding relative volume decay in the Supplementary Material. This phenomenon of aging-induced densification has been experimentally observed in protein condensates \cite{mccall_label-free_2025}.}

To determine whether the discrepancy arises from limitations in our mean field model or from the system's inability to further equilibrate, we plot in Fig.~\ref{fig:time_evolution}~\textbf{(c)} the free energy per unit length difference between our system and the theoretical free energy of an unconstrained polymer.
Calculated as $f_\text{free} = F_\text{free}/L = NE_b/L - \log (4\pi)$ (where $4\pi$ is the solid angle between two monomers).
We use a fixed lineic density of stickers ($N/L = 5.10^{-2}$) and different system sizes, characterized by $L$.
We observe a slow logarithmic decay of the free energy, followed by a plateau, where the total relaxation time depends on the system size. This type of logarithmic relaxation is usually associated with glassy systems \cite{gotzeLogarithmicRelaxationGlassforming2002}, and we propose a model in the next section to better understand this unique behavior.
Looking at the height of the plateau, we notice that it first increases with the system size before becoming independent of it.
This is an unexpected behavior as the system is self-similar along the polymer, and the equilibrium free energy per unit length is an intensive quantity with respect to $L$.
Additionally, we show in Supplementary Material that the height of the plateau also depends on the diffusion constant, $k_\text{diff}$.
These results suggest that systems above a certain size are unable to fully relax to equilibrium.
Instead, they form substructures, each equilibrated, that we call clusters.
We propose in Fig.~\ref{fig:time_evolution}~\textbf{(d)} a 1D schematic of this picture. Starting from a homogeneous distribution of stickers, they aggregate into equilibrated clusters. If the total system size is similar to the typical cluster size, the plateau free energy is the equilibrium one. 
For a system much larger than the cluster size, the plateau free energy is higher in average and intensive again.

To further examine the formation of equilibrated subsystems, we employ a hierarchical clustering algorithm to group stickers into clusters. This algorithm uses a single maximum length to define connectivity, here, we use the characteristic lengthscale of Eq.~\eqref{eq:pcf} to match the definition of clusters as equilibrated subsystems. 
Fig.~\ref{fig:cluster}~\textbf{(a)} illustrates the time evolution of the relative cluster size, showing that smaller systems tend to converge into a single cluster, while larger systems form clusters of fixed size, effectively showing a decay in the relative size. We show in Supplementary Material illustrative snapshots of the these systems.

\subsection*{Origin of the glassy relaxation}
\begin{figure}
\centering
\includegraphics[width=0.45\textwidth]{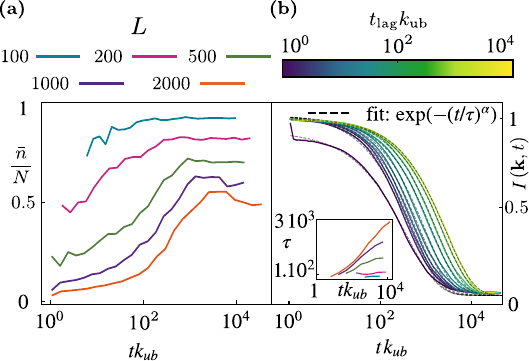}
\caption{
\textbf{(a)} Average cluster size evolution, clusters are built using a hierarchical algorithm that use a minimum distance linkage criteria. The distance used is the average sticker distance computed using the pair correlation function of Eq.~\eqref{eq:pcf}, in our case $\bar{d} = 3.6$. 
\textbf{(b)} Real part, of the intermediate scattering function for $|\bm{k}| = 1/\bar{r} = 0.37$, where $\bar{r}$ is the average distance between stickers. The stretched exponential fits are in good agreement, and the corresponding $\tau$ values obtained are plotted in the inset. The value of $\alpha$ does not evolve significantly over time, and remains $\approx 0.7$. Notice the abrut decay for small $t_\text{lag}$ that corresponds to the initial binding regime.}
\label{fig:cluster}
\end{figure}

We now investigate the origin of the slow logarithmic relaxation dynamics of the free energy. Logarithmic decay typically implies a continuous slowing down of the dynamics during relaxation. To confirm this, we examine the intermediate scattering function (ISF), defined as:
\begin{equation}
\red{
I(\bm{k},t,t_\text{lag}) = \frac{1}{N}\sum_{i=0}^N \left< e^{i\bm{k}(\bm{r}_i(t) - \bm{r}_i(t_\text{lag}))} \right>.}
\end{equation}
\red{In an isotropic system, the imaginary part vanishes, and the real part}
of the ISF characterizes the spatial and temporal dependence of the relaxation dynamics, It is commonly used to study glassy relaxations \cite{janssenModeCouplingTheoryGlass2018}. Here, we focus on the wave vector associated with the inter-sticker distance,$k = 1/<r>$, where $<r>$ is the average sticker distance computed from Eq.~\eqref{eq:pcf}. In Fig.~\ref{fig:cluster}~\textbf{(b)}, we plot the decay of the ISF over time, quantifying the decorrelation of the stickers' positions after a lag time denoted $t_\text{lag}$ along the relaxation. Unlike an exponential decay consistent with a normal random walk, we observe a stretched exponential relaxation defined by $\exp((-t/\tau)^\alpha)$, with $\alpha \approx 0.7$. This stretched exponential decay is characteristic of the $\alpha$ relaxation in glassy liquids \cite{wuStretchedCompressedExponentials2018,kobTestingModecouplingTheory1995} and indicates a relaxation slower than exponential, usually associated with subdiffusive microscopic processes \cite{bouchaudAnomalousRelaxationComplex2008} \red{and heavy tail of the displacement probability distribution}. By fitting the measured ISF  with a stretched exponential, we extract the characteristic relaxation time of the system. The inset of Fig.~\ref{fig:cluster}~\textbf{(b)} shows the increase of the characteristic relaxation time $\tau$ as a function of the simulation time, confirming our intuition of a dynamical slow-down.

We now propose a model for the origin of this dynamical slow-down. In the previous section, we showed that equilibration can be interpreted as the growth of clusters. Individual exchanges of stickers between clusters do not modify the average size; to increase the average cluster size, it is necessary to dissolve another cluster. 
To compute the dissolution time of a cluster, we consider a system with $N$ stickers spread inside different clusters. The average number of stickers in a cluster is denoted $\bar{n}$. 
We model the exchange of stickers as a two-step process: unbinding from one cluster and then rebinding in a random cluster, selected uniformly and with equal probability from all clusters. 
Each exchange occurs according to a Poisson process with a constant rate $1/\tau_\text{exch}$, which effectively encapsulates a complex process of diffusion coupled to the binding/unbinding dynamic, that we do not investigate here. For a given cluster, \red{in between two dissolution event,} the probability $P(n,t)$ to contain $n$ stickers at time $t$ evolves as:
\begin{equation}
\begin{aligned}
\frac{\text{d}P(n,t)}{\text{d}t} = &1/\tau_\text{exch} \left[ P(n+1,t) (n+1) +P(n-1,t) \bar{n} \right. \\ & \left.- P(n,t) (n + \bar{n})\right],
\end{aligned}
\label{eq:master_eq}
\end{equation}
derived from the master equation, where a cluster with $n$ stickers loses one at a rate $1/\tau_\text{exch}n$ and gains one from any of the $N/\bar{n}$ clusters at a rate $N/(N/\bar{n})/\tau_\text{exch} = \bar{n}/\tau_\text{exch}$. 
\red{Whenever a cluster size reach $n=0$ it's disolved and $\bar{n}$ needs to be updated. Therefore, we set $n=0$ as an absorbing boundary condition:}
\begin{equation}
\begin{aligned}
    &\red{\frac{\text{d}P(0,t)}{\text{d}t} = 1/\tau_\text{exch} P(1)} \\
&\red{\frac{\text{d}P(1,t)}{\text{d}t} = 1/\tau_\text{exch} \left[ 2P(2,t) - P(1,t)(1+\bar{n})\right]}.
\end{aligned}
\end{equation}
\red{This dynamics maps onto an M/M/$\infty$ queueing system. An exact solution for the mean escape time (corresponding to the mean busy period in queueing theory) is provided in \cite{guillemin_transient_1995}. Adapting their result (Eq. 2.8) to our problem yields the following asymptotic behavior:}
\begin{equation}
\red{
\tau(\bar{n}) \approx \tau_\text{exch} \frac{e^{\bar{n}}}{\bar{n}},}
\label{eq:escape_time}
\end{equation}
which shows a rapid, exponential growth of the dissolution time with the average cluster size that grows over time.

To model the long-time collective growth of clusters resulting from the dissolution of other clusters, we assume that clusters dissolve at a rate $1/\tau(\bar{n})$, and its sticker is redistributed among the $N/\bar{n} - 1$ other clusters.
We describe the dissolution events as continuous in time, and write a differential equation for the time evolution of the average cluster size:

\begin{equation}
\frac{\text{d}\bar{n}}{\text{d}t} = 1/\tau_\text{exch} e^{-\bar{n}} \frac{\bar{n}}{N/\bar{n} - 1},
\label{eq:cluster_size}
\end{equation} 

\red{
This non-linear differential equation cannot be solved analytically; however, we show in the Supplementary Material that for large systems ($1 \ll \bar{n} \ll N$), it yields the cluster dissolution time to grow over time as $\tau(t)\sim t\log(t)$.
This corresponds to an essentially linear aging regime with a logarithmic correction, usually called "super-aging".
Fig.~\ref{fig:cluster}~\textbf{(b)} successfully capture the linear part of this scaling, 
confirming the relation between the macroscopic relaxation of the ISF and the microscopic dissolution of growing clusters. 
Our focus here is on these dominant scaling factors, neglecting secondary processes, as we believe this qualitative linear aging behavior is the most robust feature of the entropic clustering mechanism.}

\subsection*{Viscoelastic response}
\begin{figure}
    \centering
    \includegraphics{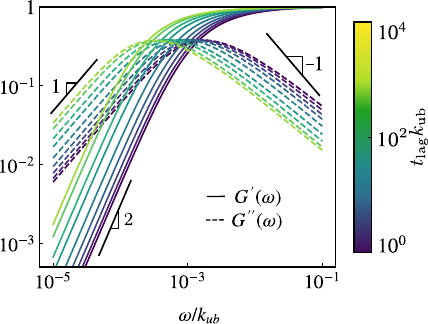}
    \caption{Estimated storage ($G^\prime$) and loss ($G^{\prime\prime}$) modulus, obtained from Eq.~\ref{eq:modulus}, using ISF measured after different lag-time. Slopes characteristics of Maxwell viscoelastic gel are displayed as reference.}
    \label{fig:modulus}
\end{figure}

{\color{black}
We now use our measurements of the intermediate scattering function (ISF) to estimate the long-time evolution of the dynamic relaxation modulus of our system.
The mechanical relaxation of a chemically crosslinked polymer gel occurs through the unbinding of stickers. When a loaded bond breaks, the stress stored in that cross-link is relieved. If the unbound sticker has sufficient mobility, future rebinding occurs without reintroducing stress into the network. In our model, bound stickers remain fixed, while unbound stickers diffuse. Consistent with our previous assumptions, we assume that unbound stickers diffuse rapidly compared to the relaxation dynamics of the polymer network.

Based on these principles, we derive in the Supplementary Material a relation connecting the ISF to the dynamic modulus for wave vector $k = 1/\langle r \rangle$: 
\begin{equation} 
\red{\label{eq:modulus} G(\omega;t_\text{lag}) = j \omega \int_{0}^{+\infty} e^{-j\omega t} I(t,t_\text{lag}) ,\text{d}t,}
\end{equation}
where $\Re[G(\omega;t_\text{lag})] = G^\prime(\omega;t_\text{lag})$ is the storage modulus, and $\Im[G(\omega;t_\text{lag})] = G^{\prime\prime}(\omega;t_\text{lag})$ is the loss modulus at angular frequency $\omega$.


The results are shown in Fig.~\ref{fig:modulus} and are consistent with previous experimental \cite{jawerthProteinCondensatesAging2020} and simulation results \cite{biswasMolecularDriversAging2024}. As expected from the growing relaxation timescales of the ISF, the crossover frequency—where the storage modulus exceeds the loss modulus, marking the transition from liquid-like to solid-like behavior-shifts to lower values as the lag time increases. This confirms that the material becomes progressively more solid-like at long times.

To aid interpretation, we also provide reference scaling laws characteristic of Maxwell viscoelastic behavior. Although the measured curves resemble those of a Maxwell material, the scalings differ slightly. This deviation arises because the ISF follows a stretched exponential form rather than a single exponential. However, the stretching exponent $\alpha$ remains close to 1, leading to a viscoelastic response that closely approximates Maxwellian behavior. 
}
\section*{Discussion}

We have introduced a minimal model to describe the dynamics of protein or RNA/DNA-protein condensates. By utilizing the general sticker and spacer framework, we examined the long-time structural evolution of these systems. Our investigation primarily focused on two general effects: the entropic cost associated with binding and the increasing difficulty of the system to rearrange upon equilibration.
Our model reveals two distinct mechanisms: the emergence of a Casimir attractive force between stickers mediated by spacers and the clustering of stickers leading to glassy relaxation. The first mechanism is reminiscent of bridging-induced phase separation, where RNA or DNA-protein binding is driven by a similar entropic force, triggering condensation \cite{ryuBridginginducedPhaseSeparation2021,brackleyNonspecificBridginginducedAttraction2013}.
\red{Here, the entropy is the one of Gaussian polymers. This is an overestimation of the reality wherein excluded volume reduces the loss of entropy due to binding.}
In our framework, this entropic attractive force appears for high binding energy, as the simultaneous unbinding of multiple stickers would disrupt the attractive interaction.
This approximation might not be universally valid for all biological systems. in fact, we anticipate that in many cases, the Casimir force will not be the primary driver of structural reorganization.
Nonetheless, in our model, this attractive force is only useful to drive sticker clustering, which is ultimately responsible for glassy relaxation. This clustering can also be driven by system-specific mechanisms, such as three-body stabilizing interactions or multiple $\beta$-sheet stacking, and has been observed in experimental systems \cite{darBiomolecularCondensatesForm2024,faragCondensatesFormedPrionlike2022,wuSingleFluorogenImaging2023}. In fact, such clustering is also a general behavior of random walkers in disordered media, commonly known as localization in this context \cite{compteLocalizationOnedimensionalRandom1998}.
For these reasons, we are confident that the clustering of stickers described here is a fairly universal feature, from which we have only highlighted the entropic minimization effect.

\red{Finally, our model offers several concrete predictions that can be directly tested in experiments. 
First, regarding the role of temperature, our framework distinguishes between thermodynamic and kinetic effects. Thermodynamically, a phase transition occurs at a critical temperature $T_c=E_c/k_B$ (Fig.~\ref{fig:Nlinkers}). Above this threshold, entropy dominates, preventing stable cross-links and aging. Below $T_c$, the system enters the aging regime where kinetics dominate: the global reorganization timescale $k_\text{ub}^{-1} \propto e^{E_b/k_B T}$. Consequently, we predict that lowering the temperature (while staying below $T_c$) will exponentially slow down the aging process. This relationship between interaction strength and relaxation dynamics is a hallmark of glassy dynamics \cite{sollichRheologySoftGlassy1997} and has been substantiated in recent protein condensate experiments \cite{alshareedahSequencespecificInteractionsDetermine2024}. 
Second, our model reconciles the seemingly contradictory aging behaviors of stiffness versus viscosity reported in \cite{jawerthProteinCondensatesAging2020}.
Using rubber elasticity theory to estimate the elastic modulus, we find $G_0= 3k_B T v_s$, where $v_s$ is the density of elastically active polymer strands. Up to a change in volume of the condensate, $v_s$ and thus $G_0$ both remain constant through aging.
On the other hand, the viscosity $\eta \approx \tau G_0$ increases with relaxation time of the growing clusters as shown in the inset of Fig.~\ref{fig:cluster}~(b).
Third regarding microrheology, the stretched-exponential decay observed in the ISF (Fig.~\ref{fig:cluster}) implies a deviation of the particle displacement probability distribution from simple Brownian motion. Consequently, we predict that this distribution will be non-Gaussian with heavy tails, exhibiting exponential rather than Gaussian decay.
Fourth, at the macroscopic level, our model predicts a gradual densification of the condensate 
driven by the reduction of the average inter-sticker distance (Fig.~\ref{fig:time_evolution}). 
We estimate the resulting relative volume contraction in the Supplementary Material (Fig. S1). 
This aging-induced shrinkage agrees with recent experimental measurements \cite{mccall_label-free_2025}.}

Our model shares phenomenological similarities with previously studied random trap models \cite{bouchaudWeakErgodicityBreaking1992,bertinSubdiffusionLocalizationOnedimensional2003}, which reproduce key features such as glass transition, subdiffusion, and localization.
However, these models typically rely on an \textit{a priori} broad (e.g., exponential) distribution of binding energies, where aging arises from the progressive exploration of ever-deeper traps.
\red{While real proteins do exhibit sequence-specific heterogeneity \cite{morganGlassyDynamicsMemory2020}, the fundamental non-covalent interactions (electrostatic, hydrophobic, hydrogen bonding) generally fall within a restricted energy range ($\approx 5-15 k_B T$). Consequently, the intrinsic energetic diversity of individual bonds is likely too narrow to justify \textit{a priori} the existence of the extremely deep traps required for long-time aging.
Our model offers a microscopic rationale for the emergence of such a broad distribution of relaxation times from a single energy scale, as proposed phenomenologically in previous studies \cite{takakiTheoryRheologyAging2023,sollichRheologySoftGlassy1997}.}
By focusing on the exponential scaling of the dissolution time in Eq.~\eqref{eq:escape_time}, we identify that a cluster of size $\bar{n}$ behaves effectively as a trap of depth $\bar{n}$ times the single-bond energy. 
This renormalization of the trap depth explains how entropic clustering can amplify small energetic heterogeneities, leading to complex viscoelastic relaxation \cite{leroyValenceCanControl2024}.

\begin{acknowledgements}
We thank Ned Wingreen for insightful discussions that contributed to improve this work and Lara Koehler for her thorough and critical reading of the manuscript. This research was supported by the Swiss National Science Foundation (grant CRSII5 193740).
\end{acknowledgements}
\section*{Data Availability}
All the data presented in this article come from c++ and python simulation. The source codes are available at :
\url{https://github.com/HugoLeRoy94/cpp_file_aging_condensate.git}
and
\url{https://github.com/HugoLeRoy94/Parallel_gillespie.git}
\bibliography{Aging_condensates}

@article{polandPhaseTransitionsOne1966,
  title = {Phase {{Transitions}} in {{One Dimension}} and the {{Helix}}---{{Coil Transition}} in {{Polyamino Acids}}},
  author = {Poland, Douglas and Scheraga, Harold A.},
  year = {1966},
  month = sep,
  journal = {The Journal of Chemical Physics},
  volume = {45},
  number = {5},
  pages = {1456--1463},
  issn = {0021-9606},
  doi = {10.1063/1.1727785},
  urldate = {2024-07-24}
}

@article{berthierDynamicHeterogeneityAmorphous2011,
  title = {Dynamic {{Heterogeneity}} in {{Amorphous Materials}}},
  author = {Berthier, Ludovic},
  year = {2011},
  month = may,
  journal = {Physics},
  volume = {4},
  pages = {42},
  publisher = {American Physical Society},
  urldate = {2024-07-24},
  copyright = {{\copyright}2011 by the American Physical Society. All rights reserved.},
  langid = {english}
}

@article{albertoparadaIdealReversiblePolymer2018,
  title = {Ideal Reversible Polymer Networks},
  author = {Alberto~Parada, German and Zhao, Xuanhe},
  year = {2018},
  journal = {Soft Matter},
  volume = {14},
  number = {25},
  pages = {5186--5196},
  publisher = {{Royal Society of Chemistry}},
  doi = {10.1039/C8SM00646F},
  urldate = {2022-06-08},
  langid = {english}
}

@article{alshareedahProgrammableViscoelasticityProteinRNA2021,
  title = {Programmable Viscoelasticity in Protein-{{RNA}} Condensates with Disordered Sticker-Spacer Polypeptides},
  author = {Alshareedah, Ibraheem and Moosa, Mahdi Muhammad and Pham, Matthew and Potoyan, Davit A. and Banerjee, Priya R.},
  year = {2021},
  month = nov,
  journal = {Nature Communications},
  volume = {12},
  number = {1},
  pages = {6620},
  publisher = {{Nature Publishing Group}},
  issn = {2041-1723},
  doi = {10.1038/s41467-021-26733-7},
  urldate = {2021-12-01},
  copyright = {2021 The Author(s)},
  langid = {english}
}

@article{bananiBiomolecularCondensatesOrganizers2017,
  title = {Biomolecular Condensates: Organizers of Cellular Biochemistry},
  shorttitle = {Biomolecular Condensates},
  author = {Banani, Salman F. and Lee, Hyun O. and Hyman, Anthony A. and Rosen, Michael K.},
  year = {2017},
  month = may,
  journal = {Nature Reviews Molecular Cell Biology},
  volume = {18},
  number = {5},
  pages = {285--298},
  publisher = {{Nature Publishing Group}},
  issn = {1471-0080},
  doi = {10.1038/nrm.2017.7},
  urldate = {2023-02-02},
  copyright = {2017 Nature Publishing Group, a division of Macmillan Publishers Limited. All Rights Reserved.},
  langid = {english}
}

@article{brackleyNonspecificBridginginducedAttraction2013,
  title = {Nonspecific Bridging-Induced Attraction Drives Clustering of {{DNA-binding}} Proteins and Genome Organization},
  author = {Brackley, Chris A. and Taylor, Stephen and Papantonis, Argyris and Cook, Peter R. and Marenduzzo, Davide},
  year = {2013},
  month = sep,
  journal = {Proceedings of the National Academy of Sciences},
  volume = {110},
  number = {38},
  pages = {E3605-E3611},
  publisher = {{Proceedings of the National Academy of Sciences}},
  doi = {10.1073/pnas.1302950110},
  urldate = {2023-04-13}
}

@article{choiPhysicalPrinciplesUnderlying2020,
  title = {Physical {{Principles Underlying}} the {{Complex Biology}} of {{Intracellular Phase Transitions}}},
  author = {Choi, Jeong-Mo and Holehouse, Alex S. and Pappu, Rohit V.},
  year = {2020},
  journal = {Annual Review of Biophysics},
  volume = {49},
  number = {1},
  pages = {107--133},
  doi = {10.1146/annurev-biophys-121219-081629},
  urldate = {2021-12-01},
  pmid = {32004090}
}

@article{hondeleDEADboxATPasesAre2019,
  title = {{{DEAD-box ATPases}} Are Global Regulators of Phase-Separated Organelles},
  author = {Hondele, Maria and Sachdev, Ruchika and Heinrich, Stephanie and Wang, Juan and Vallotton, Pascal and Fontoura, Beatriz M. A. and Weis, Karsten},
  year = {2019},
  month = sep,
  journal = {Nature},
  volume = {573},
  number = {7772},
  pages = {144--148},
  publisher = {{Nature Publishing Group}},
  issn = {1476-4687},
  doi = {10.1038/s41586-019-1502-y},
  urldate = {2022-01-27},
  copyright = {2019 The Author(s), under exclusive licence to Springer Nature Limited},
  langid = {english}
}

@article{janssenModeCouplingTheoryGlass2018,
  title = {Mode-{{Coupling Theory}} of the {{Glass Transition}}: {{A Primer}}},
  shorttitle = {Mode-{{Coupling Theory}} of the {{Glass Transition}}},
  author = {Janssen, Liesbeth M. C.},
  year = {2018},
  journal = {Frontiers in Physics},
  volume = {6},
  issn = {2296-424X},
  urldate = {2024-01-16}
}

@article{jawerthProteinCondensatesAging2020,
  title = {Protein Condensates as Aging {{Maxwell}} Fluids},
  author = {Jawerth, Louise and {Fischer-Friedrich}, Elisabeth and Saha, Suropriya and Wang, Jie and Franzmann, Titus and Zhang, Xiaojie and Sachweh, Jenny and Ruer, Martine and Ijavi, Mahdiye and Saha, Shambaditya and Mahamid, Julia and Hyman, Anthony A. and J{\"u}licher, Frank},
  year = {2020},
  month = dec,
  journal = {Science},
  volume = {370},
  number = {6522},
  pages = {1317--1323},
  publisher = {{American Association for the Advancement of Science}},
  doi = {10.1126/science.aaw4951},
  urldate = {2021-12-07}
}

@article{jawerthSaltDependentRheologySurface2018,
  title = {Salt-{{Dependent Rheology}} and {{Surface Tension}} of {{Protein Condensates Using Optical Traps}}},
  author = {Jawerth, Louise M. and Ijavi, Mahdiye and Ruer, Martine and Saha, Shambaditya and Jahnel, Marcus and Hyman, Anthony A. and J{\"u}licher, Frank and {Fischer-Friedrich}, Elisabeth},
  year = {2018},
  month = dec,
  journal = {Physical Review Letters},
  volume = {121},
  number = {25},
  pages = {258101},
  publisher = {{American Physical Society}},
  doi = {10.1103/PhysRevLett.121.258101},
  urldate = {2023-06-27}
}

@article{leeRecruitmentMRNAsGranules2020,
  title = {Recruitment of {{mRNAs}} to {{P}} Granules by Condensation with Intrinsically-Disordered Proteins},
  author = {Lee, Chih-Yung S and Putnam, Andrea and Lu, Tu and He, ShuaiXin and Ouyang, John Paul T and Seydoux, Geraldine},
  editor = {Manley, James L and Nilsen, Timothy W and Parker, Roy and Updike, Dustin L},
  year = {2020},
  month = jan,
  journal = {eLife},
  volume = {9},
  pages = {e52896},
  publisher = {{eLife Sciences Publications, Ltd}},
  issn = {2050-084X},
  doi = {10.7554/eLife.52896},
  urldate = {2021-12-03}
}

@article{linsenmeierDynamicArrestAging2022,
  title = {Dynamic Arrest and Aging of Biomolecular Condensates Are Modulated by Low-Complexity Domains, {{RNA}} and Biochemical Activity},
  author = {Linsenmeier, Miriam and Hondele, Maria and Grigolato, Fulvio and Secchi, Eleonora and Weis, Karsten and Arosio, Paolo},
  year = {2022},
  month = may,
  journal = {Nature Communications},
  volume = {13},
  number = {1},
  pages = {3030},
  publisher = {{Nature Publishing Group}},
  issn = {2041-1723},
  doi = {10.1038/s41467-022-30521-2},
  urldate = {2022-06-08},
  copyright = {2022 The Author(s)},
  langid = {english}
}

@article{morganGlassyDynamicsMemory2020,
  title = {Glassy {{Dynamics}} and {{Memory Effects}} in an {{Intrinsically Disordered Protein Construct}}},
  author = {Morgan, Ian L. and Avinery, Ram and Rahamim, Gil and Beck, Roy and Saleh, Omar A.},
  year = {2020},
  month = jul,
  journal = {Physical Review Letters},
  volume = {125},
  number = {5},
  pages = {058001},
  publisher = {{American Physical Society}},
  doi = {10.1103/PhysRevLett.125.058001},
  urldate = {2022-03-07}
}

@article{ryuBridginginducedPhaseSeparation2021,
  title = {Bridging-Induced Phase Separation Induced by Cohesin {{SMC}} Protein Complexes},
  author = {Ryu, Je-Kyung and Bouchoux, C{\'e}line and Liu, Hon Wing and Kim, Eugene and Minamino, Masashi and {de Groot}, Ralph and Katan, Allard J. and Bonato, Andrea and Marenduzzo, Davide and Michieletto, Davide and Uhlmann, Frank and Dekker, Cees},
  year = {2021},
  month = feb,
  journal = {Science Advances},
  volume = {7},
  number = {7},
  pages = {eabe5905},
  issn = {2375-2548},
  doi = {10.1126/sciadv.abe5905},
  langid = {english},
  pmcid = {PMC7875533},
  pmid = {33568486}
}

@article{takakiTheoryRheologyAging2023,
  title = {Theory of {{Rheology}} and {{Aging}} of {{Protein Condensates}}},
  author = {Takaki, Ryota and Jawerth, Louise and Popovi{\'c}, Marko and J{\"u}licher, Frank},
  year = {2023},
  month = aug,
  journal = {PRX Life},
  volume = {1},
  number = {1},
  pages = {013006},
  publisher = {{American Physical Society}},
  doi = {10.1103/PRXLife.1.013006},
  urldate = {2024-01-17}
}

@misc{wuSingleFluorogenImaging2023,
  title = {Single Fluorogen Imaging Reveals Spatial Inhomogeneities within Biomolecular Condensates},
  author = {Wu, Tingting and King, Matthew R. and Farag, Mina and Pappu, Rohit V. and Lew, Matthew D.},
  year = {2023},
  month = jan,
  primaryclass = {New Results},
  pages = {2023.01.26.525727},
  publisher = {{bioRxiv}},
  doi = {10.1101/2023.01.26.525727},
  urldate = {2023-04-20},
  archiveprefix = {bioRxiv},
  chapter = {New Results},
  copyright = {{\textcopyright} 2023, Posted by Cold Spring Harbor Laboratory. The copyright holder for this pre-print is the author. All rights reserved. The material may not be redistributed, re-used or adapted without the author's permission.},
  langid = {english}
}

@article{aulasStressspecificDifferencesAssembly2017,
  title = {Stress-Specific Differences in Assembly and Composition of Stress Granules and Related Foci},
  author = {Aulas, Ana{\"i}s and Fay, Marta M. and Lyons, Shawn M. and Achorn, Christopher A. and Kedersha, Nancy and Anderson, Paul and Ivanov, Pavel},
  year = {2017},
  month = mar,
  journal = {Journal of Cell Science},
  volume = {130},
  number = {5},
  pages = {927--937},
  issn = {0021-9533},
  doi = {10.1242/jcs.199240},
  urldate = {2024-06-13}
}

@article{bertinSubdiffusionLocalizationOnedimensional2003,
  title = {Subdiffusion and Localization in the One-Dimensional Trap Model},
  author = {Bertin, E. M. and Bouchaud, J.-P.},
  year = {2003},
  month = feb,
  journal = {Phys. Rev. E},
  volume = {67},
  number = {2},
  pages = {026128},
  publisher = {American Physical Society},
  doi = {10.1103/PhysRevE.67.026128},
  urldate = {2022-02-23}
}

@article{boeynaemsProteinPhaseSeparation2018,
  title = {Protein {{Phase Separation}}: {{A New Phase}} in {{Cell Biology}}},
  shorttitle = {Protein {{Phase Separation}}},
  author = {Boeynaems, Steven and Alberti, Simon and Fawzi, Nicolas L. and Mittag, Tanja and Polymenidou, Magdalini and Rousseau, Frederic and Schymkowitz, Joost and Shorter, James and Wolozin, Benjamin and Bosch, Ludo Van Den and Tompa, Peter and Fuxreiter, Monika},
  year = {2018},
  month = jun,
  journal = {Trends in Cell Biology},
  volume = {28},
  number = {6},
  pages = {420--435},
  publisher = {Elsevier},
  issn = {0962-8924, 1879-3088},
  doi = {10.1016/j.tcb.2018.02.004},
  urldate = {2024-06-13},
  langid = {english},
  pmid = {29602697}
}

@incollection{bouchaudAnomalousRelaxationComplex2008,
  title = {Anomalous {{Relaxation}} in {{Complex Systems}}: {{From Stretched}} to {{Compressed Exponentials}}},
  shorttitle = {Anomalous {{Relaxation}} in {{Complex Systems}}},
  booktitle = {Anomalous {{Transport}}},
  author = {Bouchaud, Jean-Philippe},
  year = {2008},
  pages = {327--345},
  publisher = {John Wiley \& Sons, Ltd},
  doi = {10.1002/9783527622979.ch11},
  urldate = {2022-02-23},
  chapter = {11},
  isbn = {978-3-527-62297-9},
  langid = {english}
}

@article{compteLocalizationOnedimensionalRandom1998,
  title = {Localization in One-Dimensional Random Random Walks},
  author = {Compte, Albert and Bouchaud, Jean-Philippe},
  year = {1998},
  month = jul,
  journal = {J. Phys. A: Math. Gen.},
  volume = {31},
  number = {29},
  pages = {6113},
  issn = {0305-4470},
  doi = {10.1088/0305-4470/31/29/004},
  urldate = {2024-02-01},
  langid = {english}
}

@article{darBiomolecularCondensatesForm2024,
  title = {Biomolecular Condensates Form Spatially Inhomogeneous Network Fluids},
  author = {Dar, Furqan and Cohen, Samuel R. and Mitrea, Diana M. and Phillips, Aaron H. and Nagy, Gergely and Leite, Wellington C. and Stanley, Christopher B. and Choi, Jeong-Mo and Kriwacki, Richard W. and Pappu, Rohit V.},
  year = {2024},
  month = apr,
  journal = {Nat Commun},
  volume = {15},
  number = {1},
  pages = {3413},
  publisher = {Nature Publishing Group},
  issn = {2041-1723},
  doi = {10.1038/s41467-024-47602-z},
  urldate = {2024-06-13},
  copyright = {2024 The Author(s)},
  langid = {english}
}

@article{faragCondensatesFormedPrionlike2022,
  title = {Condensates Formed by Prion-like Low-Complexity Domains Have Small-World Network Structures and Interfaces Defined by Expanded Conformations},
  author = {Farag, Mina and Cohen, Samuel R. and Borcherds, Wade M. and Bremer, Anne and Mittag, Tanja and Pappu, Rohit V.},
  year = {2022},
  month = dec,
  journal = {Nat Commun},
  volume = {13},
  number = {1},
  pages = {7722},
  publisher = {Nature Publishing Group},
  issn = {2041-1723},
  doi = {10.1038/s41467-022-35370-7},
  urldate = {2024-06-12},
  copyright = {2022 The Author(s)},
  langid = {english}
}

@article{garaizarAgingCanTransform2022,
  title = {Aging Can Transform Single-Component Protein Condensates into Multiphase Architectures},
  author = {Garaizar, Adiran and Espinosa, Jorge R. and Joseph, Jerelle A. and Krainer, Georg and Shen, Yi and Knowles, Tuomas P.J. and {Collepardo-Guevara}, Rosana},
  year = {2022},
  month = jun,
  journal = {Proceedings of the National Academy of Sciences},
  volume = {119},
  number = {26},
  pages = {e2119800119},
  publisher = {Proceedings of the National Academy of Sciences},
  doi = {10.1073/pnas.2119800119},
  urldate = {2024-06-13}
}

@article{gotzeLogarithmicRelaxationGlassforming2002,
  title = {Logarithmic Relaxation in Glass-Forming Systems},
  author = {G{\"o}tze, W. and Sperl, M.},
  year = {2002},
  month = jul,
  journal = {Phys Rev E Stat Nonlin Soft Matter Phys},
  volume = {66},
  number = {1 Pt 1},
  pages = {011405},
  issn = {1539-3755},
  doi = {10.1103/PhysRevE.66.011405},
  langid = {english},
  pmid = {12241362}
}

@article{kobTestingModecouplingTheory1995,
  title = {Testing Mode-Coupling Theory for a Supercooled Binary {{Lennard-Jones}} Mixture. {{II}}. {{Intermediate}} Scattering Function and Dynamic Susceptibility},
  author = {Kob, Walter and Andersen, Hans C.},
  year = {1995},
  month = oct,
  journal = {Phys. Rev. E},
  volume = {52},
  number = {4},
  pages = {4134--4153},
  publisher = {American Physical Society},
  doi = {10.1103/PhysRevE.52.4134},
  urldate = {2024-06-19}
}

@article{leroyValenceCanControl2024,
  title = {Valence Can Control the Nonexponential Viscoelastic Relaxation of Multivalent Reversible Gels},
  author = {Le Roy, Hugo and Song, Jake and Lundberg, David and Zhukhovitskiy, Aleksandr V. and Johnson, Jeremiah A. and McKinley, Gareth H. and {Holten-Andersen}, Niels and Lenz, Martin},
  year = {2024},
  month = may,
  journal = {Science Advances},
  volume = {10},
  number = {20},
  pages = {eadl5056},
  publisher = {American Association for the Advancement of Science},
  doi = {10.1126/sciadv.adl5056},
  urldate = {2024-06-12}
}

@article{ranganathanPhysicsLiquidtosolidTransitions2022,
  title = {The Physics of Liquid-to-Solid Transitions in Multi-Domain Protein Condensates},
  author = {Ranganathan, Srivastav and Shakhnovich, Eugene},
  year = {2022},
  month = jul,
  journal = {Biophysical Journal},
  volume = {121},
  number = {14},
  pages = {2751--2766},
  issn = {0006-3495},
  doi = {10.1016/j.bpj.2022.06.013},
  urldate = {2024-06-13}
}

@article{rippeLiquidLiquidPhase2022,
  title = {Liquid--{{Liquid Phase Separation}} in {{Chromatin}}},
  author = {Rippe, Karsten},
  year = {2022},
  month = feb,
  journal = {Cold Spring Harb Perspect Biol},
  volume = {14},
  number = {2},
  pages = {a040683},
  publisher = {Cold Spring Harbor Lab},
  issn = {, 1943-0264},
  doi = {10.1101/cshperspect.a040683},
  urldate = {2024-06-13},
  langid = {english},
  pmid = {34127447}
}

@article{sollichRheologySoftGlassy1997,
  title = {Rheology of {{Soft Glassy Materials}}},
  author = {Sollich, Peter and Lequeux, Fran{\c c}ois and H{\'e}braud, Pascal and Cates, Michael E.},
  year = {1997},
  month = mar,
  journal = {Phys. Rev. Lett.},
  volume = {78},
  number = {10},
  pages = {2020--2023},
  publisher = {American Physical Society},
  doi = {10.1103/PhysRevLett.78.2020},
  urldate = {2022-02-23}
}

@article{wuStretchedCompressedExponentials2018,
  title = {Stretched and Compressed Exponentials in the Relaxation Dynamics of a Metallic Glass-Forming Melt},
  author = {Wu, Zhen Wei and Kob, Walter and Wang, Wei-Hua and Xu, Limei},
  year = {2018},
  month = dec,
  journal = {Nat Commun},
  volume = {9},
  number = {1},
  pages = {5334},
  publisher = {Nature Publishing Group},
  issn = {2041-1723},
  doi = {10.1038/s41467-018-07759-w},
  urldate = {2024-06-19},
  copyright = {2018 The Author(s)},
  langid = {english}
}

@article{mccall_label-free_2025,
	title = {A label-free method for measuring the composition of multicomponent biomolecular condensates},
	volume = {17},
	copyright = {2025 The Author(s)},
	issn = {1755-4349},
	url = {https://www.nature.com/articles/s41557-025-01928-3},
	doi = {10.1038/s41557-025-01928-3},
	number = {12},
	urldate = {2026-01-19},
	journal = {Nature Chemistry},
	publisher = {Nature Publishing Group},
	author = {McCall, Patrick M. and Kim, Kyoohyun and Shevchenko, Anna and Ruer-Gruss, Martine and Peychl, Jan and Guck, Jochen and Shevchenko, Andrej and Hyman, Anthony A. and Brugués, Jan},
	month = dec,
	year = {2025},
	pages = {1891--1902},
}

@article{guillemin_transient_1995,
	title = {Transient {Characteristics} of an {M}/{M}/$\infty$  {System}},
	volume = {27},
	issn = {0001-8678},
	url = {https://www.jstor.org/stable/1428137},
	doi = {10.2307/1428137},
	number = {3},
	urldate = {2026-01-21},
	journal = {Advances in Applied Probability},
	publisher = {Applied Probability Trust},
	author = {Guillemin, Fabrice and Simonian, Alain},
	year = {1995},
	pages = {862--888},
}

@article{alshareedahSequencespecificInteractionsDetermine2024,
  title = {Sequence-Specific Interactions Determine Viscoelasticity and Ageing Dynamics of Protein Condensates},
  author = {Alshareedah, Ibraheem and Borcherds, Wade M. and Cohen, Samuel R. and Singh, Anurag and Posey, Ammon E. and Farag, Mina and Bremer, Anne and Strout, Gregory W. and Tomares, Dylan T. and Pappu, Rohit V. and Mittag, Tanja and Banerjee, Priya R.},
  year = {2024},
  month = sep,
  journal = {Nature Physics},
  volume = {20},
  number = {9},
  pages = {1482--1491},
  publisher = {Nature Publishing Group},
  issn = {1745-2481},
  doi = {10.1038/s41567-024-02558-1},
  urldate = {2025-05-27},
  copyright = {2024 The Author(s), under exclusive licence to Springer Nature Limited},
  langid = {english}
}

@article{biswasMolecularDriversAging2024,
  title = {Molecular {{Drivers}} of {{Aging}} in {{Biomolecular Condensates}}: {{Desolvation}}, {{Rigidification}}, and {{Sticker Lifetimes}}},
  shorttitle = {Molecular {{Drivers}} of {{Aging}} in {{Biomolecular Condensates}}},
  author = {Biswas, Subhadip and Potoyan, Davit A.},
  year = {2024},
  month = jun,
  journal = {PRX Life},
  volume = {2},
  number = {2},
  pages = {023011},
  publisher = {American Physical Society},
  doi = {10.1103/PRXLife.2.023011},
  urldate = {2025-05-06}
}

@article{bouchaudWeakErgodicityBreaking1992,
  title = {Weak Ergodicity Breaking and Aging in Disordered Systems},
  author = {Bouchaud, J. P.},
  year = {1992},
  month = sep,
  journal = {J. Phys. I France},
  volume = {2},
  number = {9},
  pages = {1705--1713},
  publisher = {EDP Sciences},
  issn = {1155-4304, 1286-4862},
  doi = {10.1051/jp1:1992238},
  urldate = {2022-02-23},
  copyright = {Les Editions de Physique},
  langid = {english}
}

@article{lielegSlowDynamicsInternal2011,
  title = {Slow Dynamics and Internal Stress Relaxation in Bundled Cytoskeletal Networks},
  author = {Lieleg, O. and Kayser, J. and Brambilla, G. and Cipelletti, L. and Bausch, A. R.},
  year = {2011},
  month = mar,
  journal = {Nature Mater},
  volume = {10},
  number = {3},
  pages = {236--242},
  publisher = {Nature Publishing Group},
  issn = {1476-4660},
  doi = {10.1038/nmat2939},
  urldate = {2021-12-01},
  copyright = {2011 Nature Publishing Group},
  langid = {english}
}

@article{rayASynucleinAggregationNucleates2020,
  title = {{$\alpha$}-{{Synuclein}} Aggregation Nucleates through Liquid--Liquid Phase Separation},
  author = {Ray, Soumik and Singh, Nitu and Kumar, Rakesh and Patel, Komal and Pandey, Satyaprakash and Datta, Debalina and Mahato, Jaladhar and Panigrahi, Rajlaxmi and Navalkar, Ambuja and Mehra, Surabhi and Gadhe, Laxmikant and Chatterjee, Debdeep and Sawner, Ajay Singh and Maiti, Siddhartha and Bhatia, Sandhya and Gerez, Juan Atilio and Chowdhury, Arindam and Kumar, Ashutosh and Padinhateeri, Ranjith and Riek, Roland and Krishnamoorthy, G. and Maji, Samir K.},
  year = {2020},
  month = aug,
  journal = {Nat. Chem.},
  volume = {12},
  number = {8},
  pages = {705--716},
  publisher = {Nature Publishing Group},
  issn = {1755-4349},
  doi = {10.1038/s41557-020-0465-9},
  urldate = {2023-11-06},
  copyright = {2020 The Author(s), under exclusive licence to Springer Nature Limited},
  langid = {english}
}

@article{shenLiquidtosolidTransitionFUS2023,
  title = {The Liquid-to-Solid Transition of {{FUS}} Is Promoted by the Condensate Surface},
  author = {Shen, Yi and Chen, Anqi and Wang, Wenyun and Shen, Yinan and Ruggeri, Francesco Simone and Aime, Stefano and Wang, Zizhao and Qamar, Seema and Espinosa, Jorge R. and Garaizar, Adiran and {St George-Hyslop}, Peter and {Collepardo-Guevara}, Rosana and Weitz, David A. and Vigolo, Daniele and Knowles, Tuomas P. J.},
  year = {2023},
  month = aug,
  journal = {Proc Natl Acad Sci U S A},
  volume = {120},
  number = {33},
  pages = {e2301366120},
  issn = {1091-6490},
  doi = {10.1073/pnas.2301366120},
  langid = {english},
  pmcid = {PMC10438845},
  pmid = {37549257}
}

@article{shinLiquidPhaseCondensation2017,
  title = {Liquid Phase Condensation in Cell Physiology and Disease},
  author = {Shin, Yongdae and Brangwynne, Clifford P.},
  year = {2017},
  month = sep,
  journal = {Science},
  volume = {357},
  number = {6357},
  pages = {eaaf4382},
  publisher = {American Association for the Advancement of Science},
  doi = {10.1126/science.aaf4382},
  urldate = {2023-11-29}
}

@article{songNonMaxwellianViscoelasticStress2023,
  title = {Non-{{Maxwellian}} Viscoelastic Stress Relaxations in Soft Matter},
  author = {Song, Jake and {Holten-Andersen}, Niels and H.~McKinley, Gareth},
  year = {2023},
  journal = {Soft Matter},
  volume = {19},
  number = {41},
  pages = {7885--7906},
  publisher = {Royal Society of Chemistry},
  doi = {10.1039/D3SM00736G},
  urldate = {2024-04-08},
  langid = {english}
}

\end{document}


\title{Supplementary material\\
Entropic Clustering of Stickers Induces Aging in Biocondensates}

\author{Hugo Le Roy}
\email{h.leroy@epfl.ch}
\affiliation{Institute of Physics, \'Ecole Polytechnique F\'ed\'erale de Lausanne---EPFL, 1015 Lausanne, Switzerland}

\author{Paolo De Los Rios}
\affiliation{Institute of Physics, \'Ecole Polytechnique F\'ed\'erale de Lausanne---EPFL, 1015 Lausanne, Switzerland}

\maketitle
\newpage
\section{Derivation of the binding rate}
\subsection{Probability Densities of Polymers}
\subsubsection{Polymer bound at one extremity}
We first consider a Gaussian polymer of total length $\ell$ in unit of monomer length. By considering a conformation of the polymer as a random walk, we write the probability density that its two extremities are at $\bm{r}$ from each other is given by:
\begin{equation}
p_\text{Gauss}(|\bm{r}|,\ell) = \left(\frac{3}{2\pi \ell} \right)^{3/2} \exp\left(\frac{3\bm{r}^2}{2 \ell} \right) = \frac{\Omega(|\bm{r}|,\ell)}{(4\pi)^{\ell}},
\end{equation}
where $\Omega(\bm{r},\ell)$ is the number of microstates for which each end are located at a distance $|\bm{r}|$ from one each other, while $4\pi$ is the solid angle of the random walk, and thus $(4\pi)^{\ell}$ is the total number of microstate associated with a free polymer.
Now, considering a linker at a position $\bm{r}$, and a Gaussian polymer of length $L$ bound at one extremity in $\bm{r_0} = 0$. The probability density that a portion of the polymer located at a lineic distance $\ell$ from the origin of the polymer meets a sticker located in $\bm{r}$ can be written as:
\begin{equation}
\red{p(\bm{r},\ell) = \left(\frac{3}{2\pi \ell}\right)^{3/2} e^{-\frac{3\bm{r}^2}{2\ell}} = \frac{\Omega(\bm{r},\ell) }{(4\pi)^{\ell}}}
\label{eq:1extremity}
\end{equation}
We can write this probability density as a function of the entropy of the whole polymer:
\begin{equation}
\begin{aligned}
S_b(\bm{r},\ell) &= \underbrace{  \log(\Omega(\bm{r},\ell))}_\text{polymer bound at both extremities} + \underbrace{(L-\ell)\log(4\pi)}_\text{remaining part of the polymer}\\
S_{ub}(\ell) &= L \log(4\pi)
\end{aligned}
\end{equation}
Which gives:
\begin{equation}
p(\bm{r},\ell) = \exp\left(S_\text{b}(\bm{r},\ell) - S_\text{ub}(L) \right)
\end{equation}
\subsubsection{Polymer bound at both extremities}
Now, considering a polymer of length $L$ bound at both its extremities in $\bm{R_l}$ and $\bm{R_r}$. Following a similar reasoning as in the previous section, we write the probability that the polymer meets a sticker in $\bm{r}$, along $\ell$ as:
\begin{equation}
\begin{aligned}
p(\bm{r},\ell) &= \frac{\Omega(\bm{r}-\bm{R_l},\ell)\Omega(\bm{R_r}-\bm{r},L-\ell)}{\Omega(\bm{R_r} - \bm{R_l},L)} \\
&=  \red{\left(\frac{3L}{ 2\pi \ell (L-\ell)} \right)^{3/2} \exp\left[-\frac{3}{2}\left(\frac{(\bm{R_l} - \bm{r})^2}{\ell } + \frac{(\bm{r} - \bm{R_r})^2}{(L-\ell)} - \frac{(\bm{R_r} - \bm{R_l})^2}{L} \right)\right]}
\end{aligned}
\label{eq:2extremities}
\end{equation}
Which gives a similar expression for the rates, using:
\begin{equation}
\begin{aligned}
S_\text{b}(\bm{r},l) &= \log\left[ \Omega(\bm{r}-\bm{R_l},\ell)\Omega(\bm{R_r}-\bm{r},L-\ell) \right] \\
S_\text{ub}(\bm{r},l) &= \log\left[ \Omega(\bm{R_r} - \bm{R_l},L) \right]
\end{aligned}
\label{eq:entropy_diff}
\end{equation}
\red{\subsection{Relation with the binding rates}}
\red{To compute the binding rate, we compute the probability that any of the monomer in a polymer strand crosses a linker. Denoting the size of a linker as $b$, we write such a probability as:
\begin{equation}
    P_\text{meet}(\bm{r}) = \int_0^L p(\bm{r},\ell) \text{d}\ell b^2
\end{equation}
For simplicity, we consider the linkers size to be equal to a monomer size: $b=1$
}
\section{Gillespie algorithm}
{\color{black}
\subsection{Framework}
To model the dynamical evolution of a polymer interacting with reversible cross-linking particles: \textit{stickers}, we implemented a custom stochastic simulation based on the Gillespie algorithm. The system consists of a single polymer strand immersed in an effective bath of mobile stickers. These stickers can transiently bind to the polymer, introducing local conformational constraints that evolve over time.

The polymer is modeled as a flexible Gaussian chain, discretized into $N$ monomers. Stickers may bind to any of these sites. Between two bound stickers, the polymer forms \textit{tethered segments}. At both the polymer's ends, \textit{dangling segments} form.

Stickers are modeled as particles that diffuse freely when unbound and become immobilized upon binding to the polymer. This captures the physical effect of stickers becoming effectively cross-linked when attached, constraining at the same time local polymer fluctuations.

\subsection{Evolution}
A simulation start by distributing all stickers in space in an unbounded state. Then simply perform a succession of Gillespie steps. At each simulation step, a move $m_i$ between binding, unbinding, diffusing is selected with a probability:
\begin{equation}
    p(m_i) = \frac{k(m_i)}{\sum_i k(m_i)},
\end{equation}
where $k(m_i)$ is the rate of the move $m_i$, computed according to Eq.~(3) and (4) of the main text while diffusion occurs over a constant rate. To compute this probability, we first build an array storing the rates of every possible moves, a comulative rates array is then built. A random number between 0 and $\sum_i k(m_i)$ is drawn, and the smallest index of the cumulative array that is higher than this number determines the choosen move.
To compute the entropy difference of Eq.~\ref{eq:entropy_diff} required for the binding rates, we need to keep track of the length and position of each tethered segments upon successive binding.
As a consequence, the array rate, and the cumulative array are then updated according to the local changes induced by the move.
Finally, the time increment is drawn from an exponential distribution with a characteristic time equal to the inverse total rate.


\subsection{Outputs}
The output of the system consists in a time serie of positional array of all the stickers position at each timestep, and an array of timesteps. To compute ensemble average of the time evolution of the system, we simulate an ensemble of systems between $50$ to $500$ depending on the total size of the system. The average is then performed by binning the measurements into a common time grid, while weighting the average by the time spent in each state. The value of a measurement $\mathcal{M}^n$ in the bin index $n$ is then given by:
\begin{equation}
    \mathcal{M}^n = \frac{1}{\sum_i t_i(n+1) - t_i(n)}\sum_i\mathcal{M}_i[t_i(n)]  [t_i(n+1) - t_i(n)],
\end{equation}
where $t_i(n)$ is the time (or potentially an array of time) of the system $i$ that fits within the bin of index $n$, and $\mathcal{M}_i(t_i(n))$ is the measurement performed in the system $i$ at this time.
}
\section{Mean Field Pair Probability Distribution}
\subsection{Two stickers case}
Focusing on the case of two stickers, one located in $\bm{0}$ and the second in $\bm{r}$, with a single polymer in its vicinity.
\red{We first focus on the computation of the b inding rate:}
\begin{equation}
k_\text{b}(\bm{r}) = \int_0^L \text{d}\ell p_\text{b}(|\bm{r}|,\ell)/\tau_0 = \int_0^L \text{d}\ell k_\text{b}(|\bm{r}|,\ell) = \frac{3}{2 \tau_0 \pi |\bm{r}|}\text{erfc}\left(\sqrt{\frac{3}{2 L}}|\bm{r}|\right),
\end{equation}
We then define the total probability density of the linker as:
\begin{equation}
p(\bm{r}) = p_\text{ub}(\bm{r}) + p_\text{b}(\bm{r})
\end{equation}
At equilibrium, there are no diffusive, and chemical fluxes, which means that the binding / unbinding process 
\red{of Eq.~1 of the main text} can be equilibrated independently of the diffusion, which gives:
\begin{align}
p_\text{ub}(\bm{r}) &= \frac{k_\text{ub}}{k_\text{b}(\bm{r})+k_\text{ub}} p(\bm{r}) \label{eq:pubp}\\
p_\text{b}(\bm{r}) &= \frac{k_\text{b}(\bm{r})}{k_\text{b}(\bm{r})+k_\text{ub}} p(\bm{r})
\end{align}
We now sum \red{the two equations of Eq.~1 of the main text}
and replace the expression of $p_\text{ub}(\bm{r})$ of Eq.~\ref{eq:pubp} to obtain an equation on the total probability density of finding two stickers at a relative position $\bm{r}$ from one another:
\begin{equation}
\red{\nabla^2 \left[ \frac{k_\text{ub}}{k_\text{ub}+k_\text{b}(\bm{r})} p(\bm{r}) \right] = 0}
\end{equation}
Which assumed that $\partial p(\bm{r},t) = 0$. Additionally assuming no flux of probability at the boundary, we get:
\begin{equation}
\frac{k_\text{ub}}{k_\text{b}(\bm{r})+k_\text{ub}}p_\text{eq}(\bm{r}) \stackrel{k_\text{ub} \ll k_\text{b}}{\approx} \frac{k_\text{ub}}{k_\text{b}(\bm{r})}p_\text{eq}(\bm{r}) = A
\end{equation}
Where the last approximation is valid for $E_b \gg k_bT$ \red{and for $|\bm{r}|$ of order one. In practice, the probability distribution decays fast as $\bm{r}$ increases and becomes negligible when the approximation break down.}
As a consequence, we have $\forall \bm{r}$ $k_\text{b}(\bm{r}) \stackrel{r \approx 1 }{\gg} k_\text{ub}$. We thus find the value of $A$ through the following normalizing condition:
\begin{equation}
\int \text{d}\bm{r} p_\text{eq}(\bm{r}) = 1 = A \int \text{d}\bm{r}\frac{k_\text{b}(\bm{r})}{k_\text{ub}} = \frac{A}{k_\text{ub} \tau_0} \int \text{d}\ell \underbrace{\int \text{d}\bm{r} p(\bm{r},\ell)}_{=1 \text{by normalization of the Gaussian probability}} \Rightarrow A= \tau_0 k_\text{ub}
\end{equation}
Thus, we find:
\begin{equation}
p_\text{eq}(\bm{r}) = \tau_0 k_\text{b}(\bm{r})
\end{equation}
\subsection{$N$ stickers case}
To derive a similar probability distribution, we first integrate $p(\bm{r},\ell)$ over $\ell$ for the case of a polymer bound at both of its extremities. To do so, we notice that $p(\bm{r},\ell)$ from Eq.~\eqref{eq:2extremities} decay exponentially fast with the distance from each of the neighboring node. As a result, we consider that it only takes meaningful value for $\bm{r} \approx \bm{R}_l$ or $\bm{r} \approx \bm{R}_r$, provided that $|\bm{R}_l - \bm{R}_r|\gg \sqrt{\ell}$. In this case, we can neglect the change in entropy from the largest portion of the polymer, and write for the case $\bm{r} \approx \bm{R}_r$:
\begin{equation}
    p(\bm{r},\ell) \propto \Omega((\bm{r}-\bm{R}_r),l)
\end{equation}
and for $\bm{r} \approx \bm{R}_l$:
\begin{equation}
    p(\bm{r},\ell) \propto \Omega((\bm{r}-\bm{R}_l),l)
\end{equation}
Which after computing the normalization factor, and translating the reference frame to $\bm{R}_\ell=0$, gives the same formula as Eq.~\eqref{eq:1extremity}.
As a result, the attractive Casimir-like interaction between nodes becomes a simple pair interaction.
We notice that the equilibrium distance distribution between stickers involves many binding-unbinding events.
Consequently, we consider the average polymer length between two stickers to compute their interaction.
Assuming that a polymer strand is spread equally between stickers, the average length between two stickers is $L/N$.
Because the force between stickers is now a pair interaction, and because we neglected excluded volume, the distance distribution remain of the same form, and we got:
\begin{equation}
p_\text{eq}(|\bm{r}|) = \frac{3N}{2\pi L a |\bm{r}|} \text{erfc} \left(\sqrt{\frac{3N}{2 a L}}|\bm{r}|\right)
\end{equation}

\section{Volume Contraction}

\red{To estimate the relative volume contraction of a condensate, we use the pair distance probability distribution derived previously to calculate the average distance between stickers:}
\begin{equation}
\red{\langle r \rangle = \int_0^\infty r  p(r)  4\pi r^2  \text{d}r}
\end{equation}

\begin{figure}[!ht]
    \includegraphics[scale=0.8]{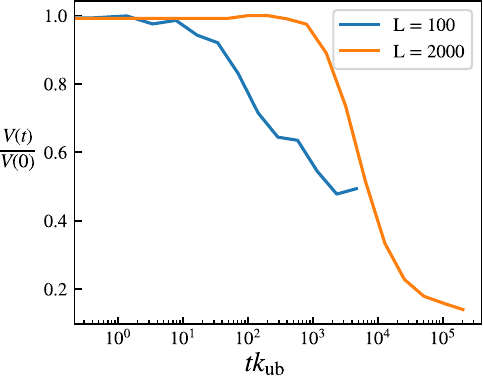}
    \caption{\label{fig:volume_contraction} Infered volume contraction obtained from the time evolution of the pair probability distribution of Fig.3 of the main text.}
\end{figure}

\section{Effect of $k_\text{diff}$ on the final state}
\begin{figure}[!ht]
\centering
\includegraphics[width=0.6\textwidth]{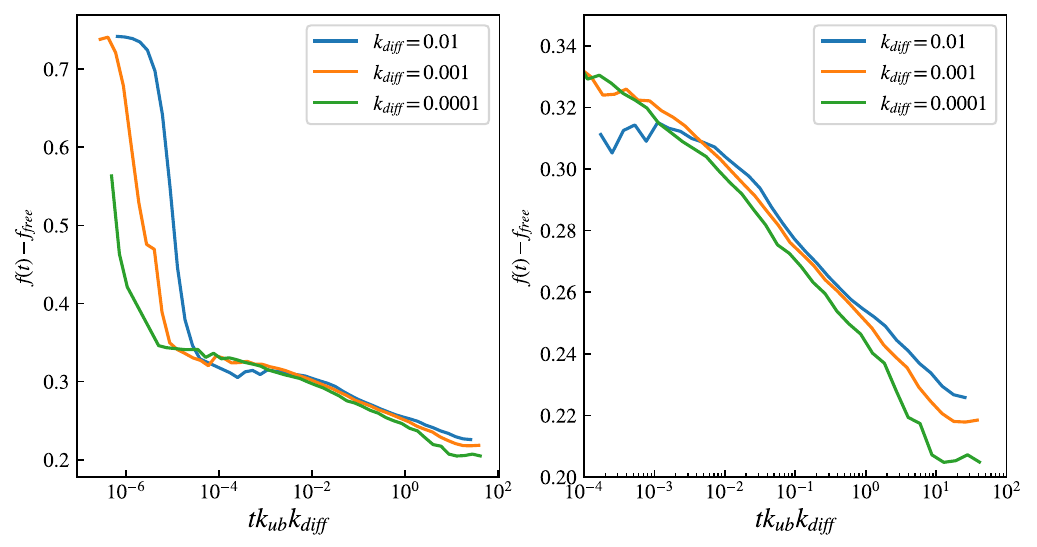}
\caption{Evolution of the lineic free energy difference as a function of time. on the left-hand side, we show the full relaxation process, including the early time behavior when stickers initially bind to the polymer. The time has been rescaled by the value of $k_\text{diff}$, consequently, the effective time of the first regime appear shifted as it does not depend on $k_\text{diff}$. On the other hand, the collapsing of the curve in the second regime highlights that the relaxation dynamic in regime is essentially proportional to the value of $k_\text{diff}$. The right-hand side graph shows a zoom on the second regime, as shown in the main text. }
\label{fig:nrg_kdiff}
\end{figure}

Although we do not explicitly study the influence of the diffusion constant on the steady state free energy, we display in Fig.~\ref{fig:nrg_kdiff} the free energy evolution for several $k_\text{diff}$.

\clearpage
\begin{figure}[!ht]
\centering
\includegraphics[width=0.9\textwidth]{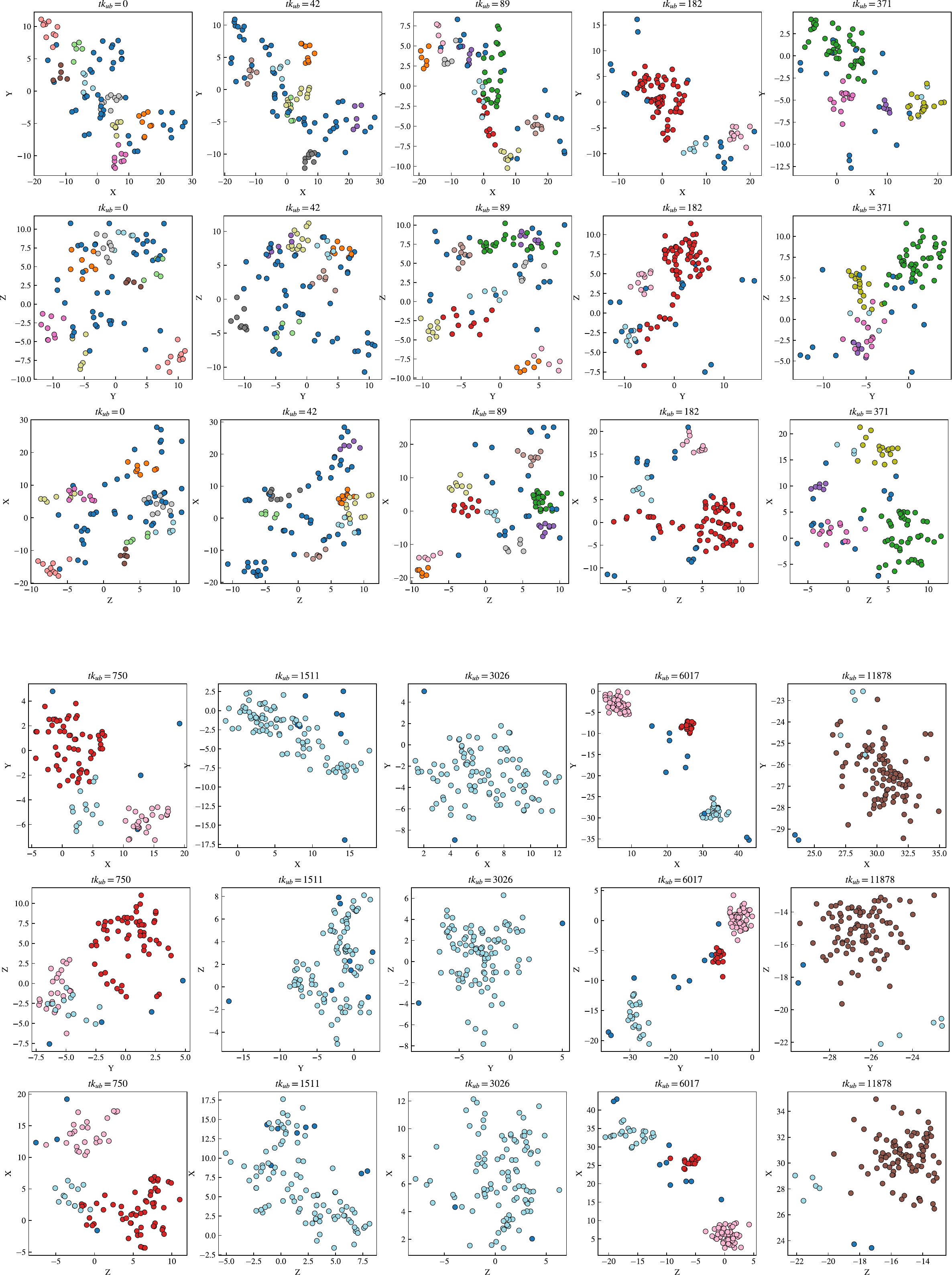}
\caption{Picture of the system, for $k_\text{diff} = 10^{-3}$, $N = 100$ stickers, and $L=2000$ monomers. The color coding correspond to the cluster labeling, each of the three lines corresponds to a different 2D projection of the 3D system.}
\label{fig:picture}
\end{figure}

\section{Logarithmic regime for cluster growth}
We start from the Eq.~(9) from the main text, and take the thermodynamic limit $N\gg 1$. Additionally, we assume that the number of remaining clusters is large: meaning that we are far from the plateau from Fig.~4~\textbf{(b)} of the main text. In this case, we have: $N/bar{n} - 1 \sim N/\bar{n}$, and the differential equation becomes:
\begin{equation}
\frac{\text{d}\bar{n}}{\text{d}t} = \frac{e^{-\bar{n}} \bar{n}^2}{\tau N}.
\end{equation}
Where we have dropped the $_\text{exch}$ for convenience.
With a change of variable : $u = t/(\tau N)$, we got:
\begin{equation}
\frac{\text{d}\bar{n}}{\text{d}u} = e^{-\bar{n}} \bar{n}^2,
\end{equation}
which we can integrate:
\begin{equation}
\int_{\bar{n}=1}^{\bar{n}(t)} \frac{e^{\bar{n}^{'}}}{\bar{n}^{'2}}\text{d}\bar{n}^{'} = \int_0^{t/(\tau N) } \text{d}u.
\end{equation}
This can be computed to obtain:
\begin{equation}
\red{
e - \text{Ei}(1) - \left[ \frac{e^{\bar{n}(t)}}{\bar{n}(t)} - \text{Ei}(\bar{n}(t))\right] = \frac{t}{\tau N},
}
\label{eq:sol:1}
\end{equation}
where $\text{Ei}$ is the exponential integral function defined as:
\begin{equation}
\text{Ei}(x) = \int_ {-\infty}^x \frac{e^t}{t}\text{d}t.
\end{equation}
From the asymptotic series of the exponential integral, we get:
\begin{equation}
\red{
\text{Ei}(x) = \frac{e^x}{x} \left(1 + \frac{1}{x} + O\left(\frac{1}{x^2}\right)\right),}
\end{equation}
thus, for $x$ large, we write the left-hand side of Eq.~\eqref{eq:sol:1} as:
\begin{equation}
\begin{aligned}
\red{e - \text{Ei}(1) - \left[ \frac{e^{\bar{n}(t)} }{\bar{n}(t)} - \text{Ei}(\bar{n}(t))\right]} &\red{= e - \text{Ei}(1) - \left[ \frac{e^{\bar{n}(t)} }{\bar{n}(t)} - \frac{e^{\bar{n}(t)}}{\bar{n}(t)} \left(1 + \frac{1}{\bar{n}(t)}+ O\left(\frac{1}{\bar{n}(t)^2}\right)\right)\right]}
\\
&\red{\underset{\bar{n}\gg1}{\sim} \frac{e^{\bar{n}(t)}}{\bar{n}(t)^2}},
\end{aligned}
\end{equation}
which we plug into Eq.~\eqref{eq:sol:1} to obtain:
\begin{equation}
\red{
\frac{e^{\bar{n}(t)}}{\bar{n}(t)^2} \sim \frac{t}{\tau N}.}
\label{eq:transcendent}
\end{equation}
For large $\bar{n}$, the left-hand side behaves like an exponential and we find:
\begin{equation}
\bar{n}(t) \propto \log(t).
\label{eq:logscale}
\end{equation}

To compute the relation with the relaxation time, we use substitute Eq.~\ref{eq:transcendent} into Eq. (9) of the main text, to find:
\begin{equation}
\tau(\bar{n}) = t\bar{n}/N
\label{eq:scaling}
\end{equation}
utilizing Eq.~\ref{eq:logscale}, we find $\tau(\bar{n}) \propto t\log(t)$.

\section{Computation of the viscoelastic modulus}

\subsection{Relation between ISF and stress relaxation}

In this section, we compute an estimation of the dynamic viscoelastic modulus of our system. To do so, we build upon the relation provided by the fluctuation response theorem between mechanical response and the system's dynamic.

In the main text, we use the intermediate scattering function to characterize the collective rearrangement of the system, defined as: 
\begin{equation}
I(\bm{k},t,t_\text{lag}) = \operatorname{Re}\left[ \left< \left< \left< e^{j\bm{k}(\bm{r}_i(t) - \bm{r}_i(t_\text{lag})} \right>_\theta \right>_i \right>_\mathcal{C}\right],
\end{equation}
Where the brackets: $<.>_\theta$, $<.>_i$, $<.>_\mathcal{C}$  respectively referes to average over the orientation of the wave vector $\textbf{k}$, over the particles in the system, and over different configuration. Looking at the first average: 

\begin{equation}
    \red{
    \left<e^{j \textbf{k}(\textbf{r}_i(t) - \textbf{r}_i(t_\text{lag}))}\right>_\theta = \frac{1}{4\pi} \int_0^{2\pi} \text{d}\phi \int_0^{\pi} e^{j k \delta r_i \cos(\theta)} \sin(\theta) \text{d}\theta = \frac{\sin(k \delta r_i)}{k \delta r_i},}
\end{equation}
where $\delta r_i(t,t_\text{lag}) = |\textbf{r}_i(t) - \textbf{r}_i(t_\text{lag})|$. 
If the sticker has remained bounded between the time $t_\text{lag}$ and $t$, $\delta r_i = 0$. On the other hand, in the paper we have assumed that diffusion is extremely fast, as a consequence, for small value of the wave vector, if the sticker unbounded $\delta r_i \gg 1/k$. With $\red{\sin(x)/x = 1}$ and $\red{\lim_{x\rightarrow \infty} \sin(x)/x = 0}$ leading to:

\begin{equation}
    \left<e^{j \textbf{k}(\textbf{r}_i(t) - \textbf{r}_i(0)}\right>_\theta \approx \left\{
    \begin{aligned}
        & 1 \text{ if $i$ remained bounded} \\
        & 0 \text{ if $i$ has unbounded},
    \end{aligned}
    \right.
\end{equation}

as a result, we can approximate the ensemble average as :
\begin{equation}
    \left< \left<e^{j \textbf{k}(\textbf{r}_i(t) - \textbf{r}_i(t_\text{lag})}\right>_\theta \right>_\mathcal{C} \approx S_i(t),
\end{equation}
where $S_i(t)$ is the survival probability of the bond $i$.

\begin{figure}
    \includegraphics[width=10cm]{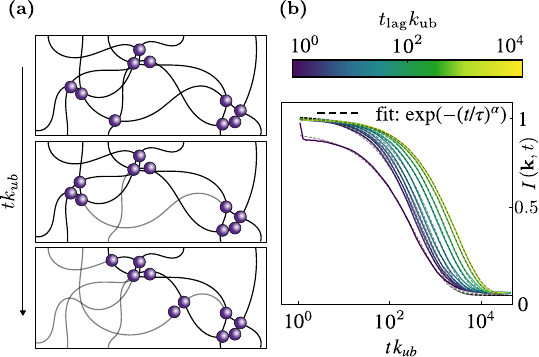}
    \caption{\label{fig:isf_stress}\textbf{(a)}: Schematic representation of stress relaxation in a heterogeneous associative polymer network. Bonds that carry stress are represented as plain black, while those in faded black represent bonds for which the stress has already been dissipated. Over time, stress relaxation occurs when a cluster is dissolved. \textbf{(b)} Modulus, of the intermediate scattering function for $|\bm{k}| = 1/\bar{r} = 0.37$, where $\bar{r}$ is the average distance between stickers. The stretched exponential fits are in good agreement, and the corresponding $\tau$ values obtained are plotted in the inset. The value of $\alpha$ does not evolve significantly over time, and remains $\approx 0.7$. Notice the abrut decay for small $t_\text{lag}$ that corresponds to the initial binding regime.}    
\end{figure}

Now, assuming that a step strain has been applied to the gel at time $t_\text{lag}$. The relaxation of the stress in such associative polymer networks occurs through a succession of unbinding and rebinding events \cite{albertoparadaIdealReversiblePolymer2018} as represented in Fig.~\ref{fig:isf_stress}. As a result, the relative decay of the the stress in the system is:
\begin{equation}
    \frac{\sigma(t)}{\sigma(t_\text{lag})} = \left< S_i(t)\right>_i \approx I(\bm{k},t,t_\text{lag}),
\end{equation}
Where $\sigma(t_\text{lag})$ and $\sigma(t)$ are respectively the initial stress in the system and the stress after a time $t$. As a result, at small wave vector, the ISF encapsulate the dynamic of the system over the relaxation of the stress. This result is a well known consequence of the fluctuation response theorem.

\subsection{Computation of the dynamic modulus}
In the linear response regime, the dynamic modulus $G(\omega; t_\text{lag})$ relates the Fourier transform of the stress to that of the strain. Since the system is aging (the cluster size distribution evolves with time), the modulus depends explicitly on the age of the system $t_\text{lag}$ when the deformation is probed.
\begin{equation}
    \sigma(\omega) = G(\omega; t_\text{lag}) \epsilon(\omega).
\end{equation}
We consider a step strain applied at time $t_\text{lag}$: $\epsilon(t) = \epsilon_0 H(t - t_\text{lag})$, where $H(t)$ is the Heaviside function. The Fourier transform of the strain is:
\begin{equation}
    \epsilon(\omega) = \int_{-\infty}^{+\infty} \epsilon_0 H(t - t_\text{lag}) e^{-j\omega t} \,\text{d}t = \frac{\epsilon_0 e^{-j\omega t_\text{lag}}}{j\omega}.
\end{equation}
In the previous subsection, we approximated the stress relaxation using the ISF and the initial stress $\sigma(t_\text{lag}) = G_0 \epsilon_0$, where $G_0$ is the instantaneous plateau modulus:
\begin{equation}
    \sigma(t) \approx \sigma(t_\text{lag}) I(k, t, t_\text{lag}) H(t - t_\text{lag}).
\end{equation}
Taking the Fourier transform of the stress:
\begin{equation}
    \sigma(\omega) = \sigma(t_\text{lag}) \int_{0}^{+\infty} I(k, t, t_\text{lag}) e^{-j\omega t} \,\text{d}t.
\end{equation}
By shifting the time variable $\tau = t - t_\text{lag}$, we obtain:
\begin{equation}
    \sigma(\omega) = \sigma(t_\text{lag}) e^{-j\omega t_\text{lag}} \int_{0}^{+\infty} I(k, \tau + t_\text{lag}, t_\text{lag}) e^{-j\omega \tau} \,\text{d}\tau.
\end{equation}
Finally, dividing $\sigma(\omega)$ by $\epsilon(\omega)$ gives the complex modulus:
\begin{equation}
    \frac{G(\omega; t_\text{lag})}{G_0} = j\omega \int_{0}^{+\infty} I(k, t+t_\text{lag}, t_\text{lag}) e^{-j\omega t} \,\text{d}t.
\end{equation}
This relation allows us to compute the frequency-dependent modulus directly from the decay of the ISF.
\bibliography{Supplement}